\documentclass[twocolumn]{openjournal}

\usepackage{amsmath, amssymb}
\usepackage{graphicx}
\usepackage{dcolumn}
\usepackage{bm}
\usepackage{xcolor}
\usepackage{aas_macros}
\usepackage{hyperref}

\begin{document}

\title{The alignment of galaxies at the Baryon Acoustic Oscillation scale}

\author{Dennis van Dompseler}
\author{Christos Georgiou}
  \email{c.georgiou@uu.nl}
\author{Nora Elisa Chisari}
  \email{n.e.chisari@uu.nl}
\affiliation{Institute for Theoretical Physics, Utrecht University, Princetonplein 5, 3584 CC, Utrecht, The Netherlands.}

\begin{abstract}
Massive elliptical galaxies align pointing their major axis towards each other in the structure of the Universe. Such alignments are well-described at large scales through a linear relation with respect to the tidal field of the large-scale structure. At such scales, galaxy alignments are sensitive to the presence of baryon acoustic oscillations (BAO). The shape of the BAO feature in galaxy alignment correlations differs from the traditional peak in the clustering correlation function. Instead, it appears as a trough feature at the BAO scale. In this work, we show that this feature can be explained by a simple toy model of tidal fields from a spherical shell of matter. This helps give a physical insight for the feature and highlights the need for tailored template-based identification methods for the BAO in alignment statistics. We also discuss the impact of projection baselines and photometric redshift uncertainties for identifying the BAO in intrinsic alignment measurements.
\end{abstract}

\maketitle

\section{\label{sec:intro}Introduction}

Baryon acoustic oscillations (BAO, \citep{Bassett10}) are sound waves supported by the plasma present in the Universe before recombination. After the Universe became neutral, these waves could no longer travel and remained frozen at a comoving scale of $\sim 150$ Mpc. In the late Universe, BAO manifest themselves as a subtle but significant percent-level peak in the auto-correlation function of galaxies or matter. Because they constitute a standard ruler of an absolute distance scale, they are regularly used to probe the expansion of the Universe \citep{Weinberg13}.
   
Any cosmological observable that correlates with the matter field can have a manifestation of BAO. One such observable beyond galaxy clustering statistics is the alignments of galaxies. Elliptical galaxies are known to align their major axis radially towards other galaxies \citep{Brown02}, and this phenomenon can be described, when the alignment is weak, by a proportional response of the projected shape of a galaxy to the projected tidal field of matter \citep{Catelan01}. This model is successful in describing the observed alignments of luminous red galaxies at large-scale from wide surveys \citep{Blazek11,Joachimi11,Singh15,Johnston18,Fortuna21}. Although intrinsic alignments are typically regarded a contaminant to other cosmological observables \citep{Brown02,Hirata09,Kirk12,Krause16,Zwetsloot}, there are examples of how they can be used for extracting cosmological information \citep{CD13,Chisari14,SCD15,Biagetti20,Taruya20}.

In principle, a detection of BAO could be achieved in the correlation function of galaxy alignments around other galaxies. In \citep{CD13}, it was shown that such a detection was within the reach of existing surveys. For luminous galaxies in the Baryon Oscillation Spectroscopic Survey (BOSS, \citep{Dawson13}), the signal-to-noise ratio ($S/N$) would be of the order of $\sim 2.7$. For upcoming data sets such as the Dark Energy Spectroscopic Instrument (DESI, \citep{DESI}), the expectation is for this to increase to $S/N\sim 12$.

Searches for BAO in galaxy statistics often adopt matched templates \citep{Seo07,Seo10}, decompositions thereof \citep{wavelet,wavelet2} or remove the smooth (no BAO) component \citep{Percival07}. In \citep{CD13}, it was noticed that the shape of the BAO differs from the traditionally expected `peak' at $150$ Mpc. When looking at the alignment of galaxies with the matter field, it rather appears as a trough at a similar distance, followed by a peak at larger comoving separations. This behavior was recently confirmed by \citep{Okumura19}, who measured the alignment of massive (cluster-scale) halos with the underlying matter field in the {\sc DarkQuest} N-body simulations \citep{DarkQuest}. These authors also pointed out a similar behaviour for the correlation of halo alignments with the velocity field, with the BAO appearing as trough rather than a peak.

In light of possible upcoming detections of this feature, we aim here to give an intuitive physical picture of the origin of this trough pattern rooted in simple linear physics. We show that gravitational tides in and around a spherical shell of matter display exactly the trough pattern and justify its appearance in both matter- and velocity-alignment cross-correlations. We also discuss the impact of long projection baselines and photometric redshifts for identifying the BAO in observational data.

This work is organised as follows. In Section \ref{sec:model}, we introduce the most widely used linear model for the shapes of galaxies and halos, we present the equations for correlations with matter, galaxies and velocity field, we explain how we model BAO, and how tidal fields are calculated for the simple toy model. Section \ref{sec:results} gives our results and we conclude in Section \ref{sec:conclusions}. 

In this work, we model a Universe with and without BAO `wiggles' using the analytical approximation of \citep{EHu} for a cosmology with $\sigma_8 = 0.8158$, $h=0.6774$, $\Omega_{\rm m}=0.3089$, $\Omega_{\rm b}=0.0486$ and $n_s=0.9667$, consistent with constraints from the {\it Planck} satellite \citep{Planck15}. The \citet{EHu} matter power spectra at $z=0$ are output by the {\sc nbodykit} software \citep{nbodykit}. Other cosmological quantities were obtained via the Core Cosmology Library \footnote{\url{https://github.com/LSSTDESC/CCL}} \cite{CCL}. In the following sections, we compare our predictions for the alignment correlation function for both models. In the matter power spectrum, BAOs appear as a series of successive peaks or `wiggles' at different wavenumbers. In real space, this corresponds to a peak in the three-dimensional correlation function of galaxies, at a comoving scale of $\sim 150$ Mpc, or equivalently, $\sim 100\,h^{-1}$ Mpc \citep{Komatsu09}.

\section{Modelling}
\label{sec:model}

\subsection{Linear alignment model}

In the linear alignment model \citep{Catelan01}, galaxies align their observed two-dimensional shapes proportionally to the projected tidal field of matter. This is mathematically described as:
\begin{equation}\label{eq:gamma+}
    (\gamma_+,\gamma_\times) = -\frac{C_1}{4\pi G}(\partial_x^2-\partial_y^2,2\partial_x\partial_y)\phi_p
\end{equation}
Here, $\gamma_+$ and $\gamma_\times$ are the shape perturbations in the radial/tangential direction ($\gamma_+$) and the perturbation rotated 45 degrees with respect to the radial/tangential direction  ($\gamma_\times)$. $C_1$ is an unknown proportionality constant, i.e. the alignment `bias' and $\phi_p$ is the primordial gravitational potential (i.e. at some high redshift when the galaxy was formed). This gives a prescription for connecting galaxy shapes to the underlying gravitational potential field and leaves $C_1$ as a free parameter. As a result, galaxy shapes are expected to be correlated with any observable that depends on the gravitational potential, or the matter field which sources it. 

The choice of redshift evolution is inconsequential for our work, since we mostly work at fixed redshift. There is, however, significant uncertainty over how galaxies gain and evolve their alignment over time. Our choice of adopting the primordial alignment model is justified by the findings of \citep{Camelio15}, who suggested that instantaneous alignment of galaxies over time should be ruled out based on theoretical considerations.

The linear alignment model is known to provide a good description of elliptical galaxies in both simulations \citep{Tenneti15,Chisari15,Chisari16,Hilbert17} and observations \citep{Joachimi11,Blazek11,Singh15,Johnston18,Fortuna21} and it is widely used in cosmological studies which aim to extract information from gravitational lensing \citep[e.g.][]{Hikage21,Heymans21,Secco22}. Here, intrinsic alignments act as a contaminant. 

We will not cover blue/spiral galaxies in this work, to which different models are thought to apply, namely based on tidal torque theory \citep{Porciani02,Porciani02b,Codis15}. Instead, we will assume that there is at least a sample of elliptical galaxies for which the linear alignment model is applicable. This assumption is based in ample observational evidence. The strength (or `bias') of alignment $C_1$ is constrained from observations \citep{Joachimi11,Blazek11,Singh15,Johnston18,Fortuna21} to be generally positive. In this context, this means that elliptical galaxies tend to point their major axis towards peaks in the density field. The authors of \citep{Camelio15} proposed a method for estimating $C_1$ using the stellar distribution function of elliptical galaxies. Using this method, once again one expects that $C_1>0$. However, the predicted alignment seemed to fall short of the observed one. This could be a consequence of alignments of galaxies being built-up over time, rather than instantaneously reacting to the tidal field. For our purposes, it suffices to emphasise that the sign of $C_1$ is at least observationally constrained for elliptical galaxies that are the subject of our work. A completely analogous model and arguments would apply for halos as well, where $C_1$ is also known to be positive \citep{vanUitert17}.

The most commonly measured statistic of galaxy intrinsic shapes is the projected correlation function of galaxy positions and the $+$ component of the shape, $w_{g+}(r_p)$, which is a function of the projected comoving separation between galaxies. At any given redshift, this is given by an integral along separation in comoving radial distance ($\Pi$) of the three-dimensional correlation of positions and shapes, $\xi_{g+}(r_p,\Pi,z)$:
\begin{equation}
    w_{g+}(r_p,z) = \int_{-\Pi_{\rm max}}^{\Pi_{\rm max}} d\Pi\,\xi_{g+}(r_p,\Pi,z). \label{eq:wg+}
\end{equation}
Here, $\xi_{g+}$ is defined as: 
\begin{equation}
    1 + \xi_{g+} = \langle [1+\delta_g(\mathbf{x_{1}})] \gamma_{+}(\mathbf{x_{2}}) \rangle,
\end{equation}
and $\mathbf{r}=\mathbf{x_2}-\mathbf{x_1}$. 

Because galaxy alignments only arise between galaxies that are physically close, $\Pi_{\rm max}$ is usually restricted to scales $\lesssim 100\,h^{-1}{\rm Mpc}$. This justifies assuming a separate dependence of $\xi_{g+}$ on $\Pi$ and redshift. 

In the linear alignment model, $\xi_{g+}(r_p,\Pi)$ is given by
\begin{widetext}
\begin{equation}
    \xi_{g+}(r_p,\Pi,z) = \frac{b_{\rm g}C_1\rho_{\rm crit}\Omega_{\rm m}}{2\pi^2D(z)} \int_0^\infty dk_z\,\int_0^\infty dk_\perp \,\frac{k_\perp^3}{k^2}P(k,z)J_2(k_\perp r_p)\cos(k_z\Pi)\,\label{eq:xig+LA}
\end{equation}
\end{widetext}
where $D(z)$ is the growth function, normalized to $(z+1)D(z)=1$ during matter domination, $P(k,z)$ is the matter power spectrum, $\rho_{\rm crit}$ is the critical density today and $J_2$ is the Bessel function of the first kind of order $2$. The coordinates in Fourier space are given by: $k = (k_z, k_\perp)$, $k_z$ is given along the line of sight and $k_\perp$ perpendicular to it. This results in a projected correlation function 
\begin{widetext}
\begin{equation}
    w_{\rm g+}(r_p,z) = \frac{b_{\rm g}C_1\rho_{\rm crit}\Omega_{\rm m}}{\pi^2D(z)} \int_0^\infty dk_z\,\int_0^\infty dk_\perp \,\frac{k_\perp^3}{k^2k_z}P(k,z)J_2(k_\perp r_p)\sin(k_z\Pi_{\rm max})\,\label{eq:wg+LA}
\end{equation}
\end{widetext}
Notice that when $b_{\rm g}=1$, $w_{\rm g+}$ reduces to the correlation between matter and the $+$ component of galaxy shapes, $w_{\rm m+}$. For simplicity, we will work with $w_{\rm m+}$ from here on. We also note that in the linear alignment model, $w_{\rm g\times}$ is expected to be zero due to symmetry, and we do not consider it further in this work \footnote{See \citet{Biagetti20} for an exception due to parity-breaking.}.

For comparison, the projected correlation function of the matter field, $w_{\rm mm}$, is given by
\begin{widetext}
\begin{equation}
    w_{\rm mm}(r_p,z) = \frac{1}{\pi^2} \int_0^\infty dk_z\,\int_0^\infty dk_\perp \,\frac{k_\perp}{k_z}P(k,z)J_0(k_\perp r_p)\sin(k_z\Pi_{\rm max})\,\label{eq:wmm}.
\end{equation}
\end{widetext}

This model was used in forecasts by \citep{CD13}, where it was proposed that a detection of BAO could be achieved in the projected alignment correlation function of galaxies. It is also commonly used in fits to the data \citep[e.g.][]{Blazek11,Singh15}. However, other works adopt larger projections lengths, effectively taking $\Pi_{\rm max}$ to infinity \citep[e.g.][]{Johnston18}. The corresponding projected correlation functions in those cases are 
\begin{eqnarray}
    w_{\rm mm}(r_p,z) &=&  \,\int_0^\infty \frac{dk_\perp}{2\pi} \,k_\perp P(k_\perp,z)J_0(k_\perp r_p)\,\label{eq:wmm_approx},\\
    w_{\rm m+}(r_p,z) &=& \tilde C_1 \int_0^\infty \frac{dk_\perp}{2\pi} k_\perp P(k_\perp,z)J_2(k_\perp r_p)\,\label{eq:wm+_approx}.
\end{eqnarray}
where $\tilde C_1=C_1\rho_{\rm crit}\Omega_{\rm m}/D(z)$ for simplicity. Eq. \ref{eq:wm+_approx} is derived explicitly in the appendix.

Although more sensitive to shape noise, intrinsic shape auto-correlations have been derived in previous work in the context of the linear alignment model and also detected in spectroscopic survey observations \citep[e.g.][]{Blazek11}. The projected correlation functions for shape-shape correlations take the form
\begin{widetext}
\begin{equation}
    w_{(++,\times\times)}(r_p,z) = \frac{1}{2\pi^2} \left(\frac{C_1\rho_{\rm crit}\Omega_{\rm m}}{D(z)}\right)^2 \int_0^\infty dk_z\,\int_0^\infty dk_\perp \,\frac{k_\perp^5}{k^4k_z}P(k,z)[J_0(k_\perp r_p)\pm J_4(k_\perp r_p)]\sin(k_z\Pi_{\rm max}).\label{eq:wssLA}
\end{equation}
\end{widetext}

Because intrinsic alignments are correlated with the matter field, we also expect them to be correlated with the velocity field of the large-scale structure \citep{Okumura19}. On linear scales, the velocity field and the matter density are related by the continuity equation:
$\nabla\cdot{\vec v} = - \delta\,{(1+z)}/H(z)f(z)$,
where $H$ is the Hubble factor, $f=d\ln D/d\ln a$ is the logarithmic growth rate and ${\vec v}$ is the irrotational velocity field. This leads to a correlation between the divergence of the velocity field and the $+$ component of galaxy shapes. In practice, one expects to actually measure the correlation between projected $+$ shapes and {\it radial} velocities (along the line-of-sight) \citep{vanGemeren21}, which in Fourier space is $v_r\propto (k_z/k)\delta/k$. The $w_{v_r+}$ correlation function is thus modelled by
\begin{widetext}
\begin{equation}
    w_{v_r+}(r_p,z) = \frac{C_1\rho_{\rm crit}\Omega_{\rm m}(1+z)}{\pi^2D(z)H(z)f(z)}\int_0^\infty dk_z\int_0^\infty dk_\perp \frac{k_\perp^3}{k^4}P(k,z)J_2(k_\perp r_p)\sin(k_z\Pi_{\rm max}).
\end{equation}
\end{widetext}
Since the radial velocity field often requires spectroscopic information to be constructed, we do not discuss the effect of photometric redshifts on the $w_{v_r+}$ correlation function (but see \citep{vanGemeren21} for an alternative approach).

\subsection{Modelling photometric redshifts}

The correlation functions presented in the sections above assume precise knowledge of the redshift information of our galaxy samples. This would be the case when data are taken from a spectroscopic survey, but such surveys require a predetermined target selection and long integration times which limit the size of galaxy samples that can be obtained. Photometric surveys can overcome this problem at the cost of significantly reduced accuracy in the determination of redshift information by using band photometry instead of spectra. 

The accuracy of the photometric redshifts depend on a number of factors, such as the signal-to-noise of the flux measurement of galaxies and the existence of a representative calibration data-set. There are several techniques that can increase the typical accuracy of photometric redshifts. These include mapping of the galaxy red sequence (limited to intrinsically red galaxies) \cite{redmagic, KiDSLRG}, using machine learning techniques with representative overlapping spectroscopic samples as training set (limited by the training set) \cite{JoachimiPZ,KiDSBrightPZ,KiDSSOMPZ} or using narrow band photometry to resolve more features in a galaxy's spectral energy distribution (which is more observationally costly compared to broad band photometry) \cite{COSMOSPZ,PAUSPZ}. In light of these techniques, it is interesting to investigate how the projected correlation functions change when the galaxy samples used are obtained through photometric data. 

To compute the projected correlation functions in this context, we model the impact of redshift uncertainty following \cite{JoachimiPZ}. The uncertainty is expressed in the probability density function $p(z|\bar{z})$, where $z,\bar{z}$ is the true and observed redshift of a galaxy, respectively. We choose to model this with a generalized Lorentzian distribution,
\begin{equation}
p(z|\bar{z}) \propto \left(1+\frac{\Delta z^2}{2as^2}\right)^{-a}\,,
\label{eq:pzbarz}
\end{equation}
where $\Delta z=(z-\bar{z})/(1+z)$ and $a,s$ are free parameters. In \cite{KiDSBright} it was shown that this distribution better describes the probability density function compared to a Gaussian one, especially the long tails away from the mean. We fix $a=2.613$ as was found in \cite{KiDSBright} and vary $s\in\{0.0035, 0.015,0.025\}$ to mimic different photometric redshift precision scenarios. The precision is commonly expressed in terms of the scaled median absolute deviation (SMAD) of $\Delta z$, given by $\hat{\sigma}_{\Delta z}=k\cdot\mathrm{MAD}$, where $k\approx1.4826$ and $p\left(\left|\Delta z\right|\leq \mathrm{MAD}\right)=1/2$ (using the fact that the median of Eq. \ref{eq:pzbarz} is at $\Delta z=0$). The SMAD is a way to quantify a standard deviation equivalent in the case where the distribution is different than a Gaussian.

Assuming that the line-of-sight separation between two galaxy pairs is small compared to the comoving radial distance of their mean redshift, we can express their true redshifts as $z_1+z_2=2z_{\rm m}$ and 
$\Pi\approx c(z_1-z_2)/H(z_\mathrm{m})$. The matter-matter projected correlation function in the presence of redshift uncertainty can be modelled by 
\begin{equation}
w_{\rm mm}^{\rm phot}(r_p, z_{\rm m}) = \int_{-\Pi_\mathrm{max}}^{\Pi_\mathrm{max}}\mathrm{d}\Pi\int_0^\infty\frac{\mathrm{d} \ell\,\ell}{2\pi}J_0(\ell\theta)C_{\rm mm}(\ell, \bar{z_1}, \bar{z_2})\,,
\label{eq:wmm_phot}
\end{equation}
where $r_p\approx\theta\chi(z_\mathrm{m})$ and $C_\mathrm{mm}$ is the matter-matter angular power spectrum, computed using $p(z_\mathrm{m}|\bar{z}_{1,2})$ for the redshift distribution of its tracers, given by
\begin{equation}
C_\mathrm{mm}=\int_0^{\chi_\mathrm{hor}}\frac{p\left(\chi|\chi(\bar{z}_1)\right)p\left(\chi|\chi(\bar{z}_2)\right)}{\chi^2}\,P\left(\frac{\ell}{\chi}, z(\chi)\right)\,,
\end{equation}
where $\chi_\mathrm{hor}$ is the comoving horizon distance. In a similar way, one can compute the projected matter-shape and shape-shape correlation functions, using the matter-intrinsic and intrinsic-intrinsic angular power spectra, $C_\mathrm{mI}=-\tilde{C}_1\,C_\mathrm{mm}$ and $C_\mathrm{II}=\tilde{C}_1^2\,C_\mathrm{mm}$. This will lead to
\begin{equation}
w_\mathrm{m+}^\mathrm{phot}(r_p, z_\mathrm{m}) = -\int_{-\Pi_\mathrm{max}}^{\Pi_\mathrm{max}}\mathrm{d}\Pi\int_0^\infty\frac{\mathrm{d} \ell\,\ell}{2\pi}J_2(\ell\theta)C_\mathrm{mI}(\ell, \bar{z_1}, \bar{z_2})\,
\label{eq:wmI_phot}
\end{equation}
and
\begin{widetext}
\begin{equation}
w_\mathrm{(++,\times\times)}^\mathrm{phot}(r_p, z_\mathrm{m}) = \int_{-\Pi_\mathrm{max}}^{\Pi_\mathrm{max}}\mathrm{d}\Pi\int_0^\infty\frac{\mathrm{d} \ell\,\ell}{2\pi}\left[J_0(\ell\theta)\pm J_4(\ell\theta)\right]C_\mathrm{II}(\ell, \bar{z_1}, \bar{z_2})\,.
\label{eq:wII_phot}
\end{equation}
\end{widetext}

\subsection{Tidal field for a spherical mass distribution: a simple BAO model}

To give a qualitative explanation of how the BAO features in the $w_{\rm m+}$ correlation, we recall that the gravitational potential of a spherical mass distribution is given by
\begin{equation}
    \phi(r)=-4\pi G \left[\frac{1}{r}\int_0^r dr_1\rho(r_1)r_1^2+\int_r^{\infty}dr_1\rho(r_1)r_1\right]\label{eq:phi_spherical}.
\end{equation}
where $\rho(r_1)$ is the density of matter as a function of radius. 

For an extended object, the difference between the force acting at any point and the force acting at the center of mass is the tidal force: ${\bf T} = {\bf F}(\bf x)-{\bf F}({\bf x}_{\rm CM})$. A small displacement from the center of mass gives rise to a differential change of the force of $d{\rm T}_j = \tau_{ij}dx^i$, implicitly summing over $i$ and where $\tau_{ij}=-\partial_i\partial_j\phi$ is the tidal tensor. A spherically symmetric gravitational potential originates a tidal field given by \citep{Masi07}:
\begin{eqnarray}
\tau_{rr}(r)&=&-\partial^2_r\phi(r),\\
\tau_{\theta\theta}(r)&=&\tau_{\phi\phi}(r)=-\partial_r\phi(r)/r.
\end{eqnarray}
The explicit expressions in terms of the density profile of the object are
\begin{eqnarray}
    \tau_{rr}(r) &=& 4\pi G\left[\frac{2}{r^3}\int_0^r dr_1\rho(r_1)r_1^2-\rho(r)\right],\label{eq:trgen}\\
    \tau_{\theta\theta}(r) &=& \tau_{\phi\phi}(r) = -\frac{4\pi G}{r^3}\int_0^r dr_1\rho(r_1)r_1^2,\label{eq:ttgen}
\end{eqnarray}
and this can also be expressed in terms of the mean density interior to a given radius, $\bar\rho(r)$. For example, as $\tau_{rr}(r)=4\pi G[2\bar\rho(r)/3-\rho(r)]$.

We model the BAO as a spherical shell of mass $M_{\rm BAO}$, with an inner radius $R_{\rm BAO}$, width $\Delta R$ and uniform density $\rho_{\rm BAO}$. We will neglect the smooth extended component that corresponds to the matter distribution inside and outside the shell, and focus only on how the tidal field changes when the BAO shell is added. 

\begin{figure}
    \centering
    \includegraphics[width=0.47\textwidth]{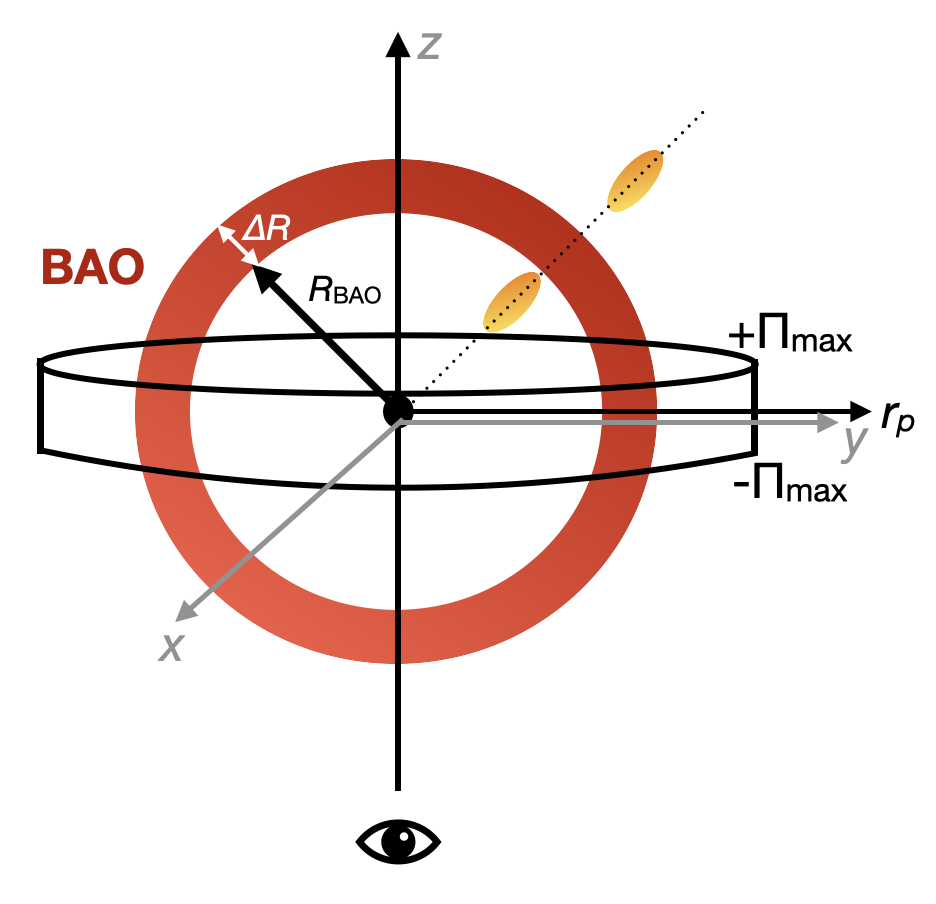}
    \caption{A sketch showing the geometry of the problem. According to observational constraints on the linear alignment model, galaxies (orange) align themselves radially towards density peaks. These constraints come from integrating the three-dimensional correlation function of galaxy positions and shapes along a line-of-sight baseline of $\Pi_{\rm max}$ (black cylinder), typically $\lesssim R_{\rm BAO}$, the BAO scale. The BAO is represented as a spherical shell of matter around the center of the potential. The reader should interpret that the smooth `no wiggles' component has been subtracted in this image.}
    \label{fig:sketch}
\end{figure}

Looking at Figure \ref{fig:sketch}, we first examine the tidal field of the mass configuration on the $xy$ plane. This should be qualitatively representative of the projection along the line of sight, although we will discuss the impact of the projection in more detail below. We imagine taking a spherical coordinate system where $\phi=0$ is aligned with the projection ($z$) axis. According to Eq. \ref{eq:gamma+}, we would then have the change in shapes due to the presence of the BAO being $\gamma_+^{\rm BAO} = C_1[\tau_{rr}(r)-\tau_{\theta\theta}(r)]/(4\pi G)$.

The radial and $\theta$ components of the tidal field for such configuration are
\begin{widetext} 
\begin{eqnarray}
\tau_{rr}(r)&=& \left\{ \begin{array}{lcrr}
         0 & & r\leq R_{\rm BAO} & {\rm (I)} \\
        -4\pi G \rho_{\rm BAO}[1/3+2/3(R_{\rm BAO}/r)^3] & & R_{\rm BAO}<r\leq R_{\rm BAO}+\Delta R & {\rm (II)} \\
        2GM_{\rm BAO}/r^3 & & r> R_{\rm BAO}+\Delta R & {\rm (III)} 
        \end{array} \label{eq:taurr} \right. \\
\nonumber\\
\nonumber\\
\tau_{\theta\theta}(r)&=& \left\{ \begin{array}{lcrr}
         0 & & r\leq R_{\rm BAO} & {\rm (I)} \\
        -4\pi G \rho_{\rm BAO}[1-(R_{\rm BAO}/r)^3]/3\qquad\qquad & & R_{\rm BAO}<r\leq R_{\rm BAO}+\Delta R & {\rm (II)} \\
        -GM_{\rm BAO}/r^3 & & r> R_{\rm BAO}+\Delta R & {\rm (III)}
        \end{array}\label{eq:tautt} \right.
\end{eqnarray}
\end{widetext}
respectively, where we have identified three regions of interest: inside the spherical shell (I), within the shell (II) and outside (III). Similarly, $\tau_{\phi\phi}(r)=\tau_{\theta\theta}(r)$. 

In addition to this simple model, we also consider a slightly more realistic Gaussian form for the density profile of the shell, with a center at $R_{\rm BAO}+\Delta R/2$ and a dispersion $\sigma_{\rm BAO}=\Delta R/2$. We obtain the tidal field in this scenario numerically integrating Eqs. \ref{eq:trgen} and \ref{eq:ttgen}. We then use the change of shapes ($\gamma_+^{\rm BAO}$) to explain deviations in $w_{\rm mm}$ from $w_{m+}$ based on the definition given in Eq. \ref{eq:wg+}.

We will also consider the effect of projection in our toy model for $\gamma_+^{\rm BAO}$ by integrating along the line of sight:
\begin{equation}
w_{\gamma_+^{\rm BAO}}(r_p)=\int_{-\Pi_{\rm max}}^{\Pi_{\rm max}} d\Pi \,\gamma_+^{\rm BAO}(\sqrt{r_p^2+\Pi^2})
\end{equation}
In practice, we perform this integration by direct summation over small $\Pi$ bins. We also considered including a dependence of the observed $\gamma_+^{\rm BAO}$ with the angle with respect to the line of sight: $\mu=\Pi/r$. This factor is expected from the linear alignment model \citep{Okumura19} but it does not change any of our results significantly at the BAO scale.

\begin{figure}
    \centering
    \includegraphics[width=0.47\textwidth]{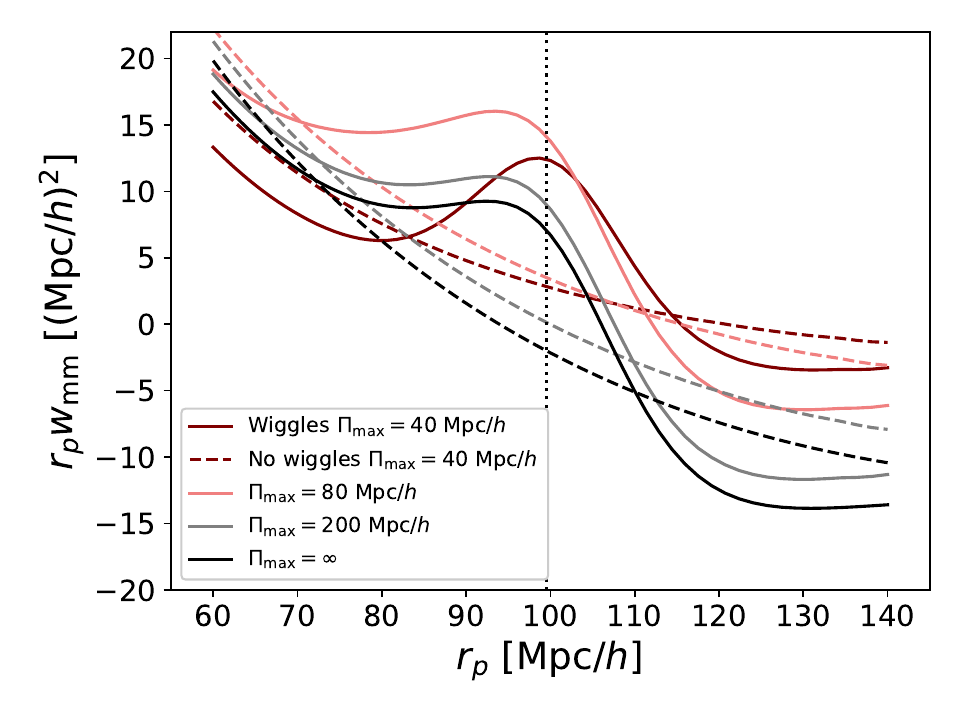}
    \includegraphics[width=0.47\textwidth]{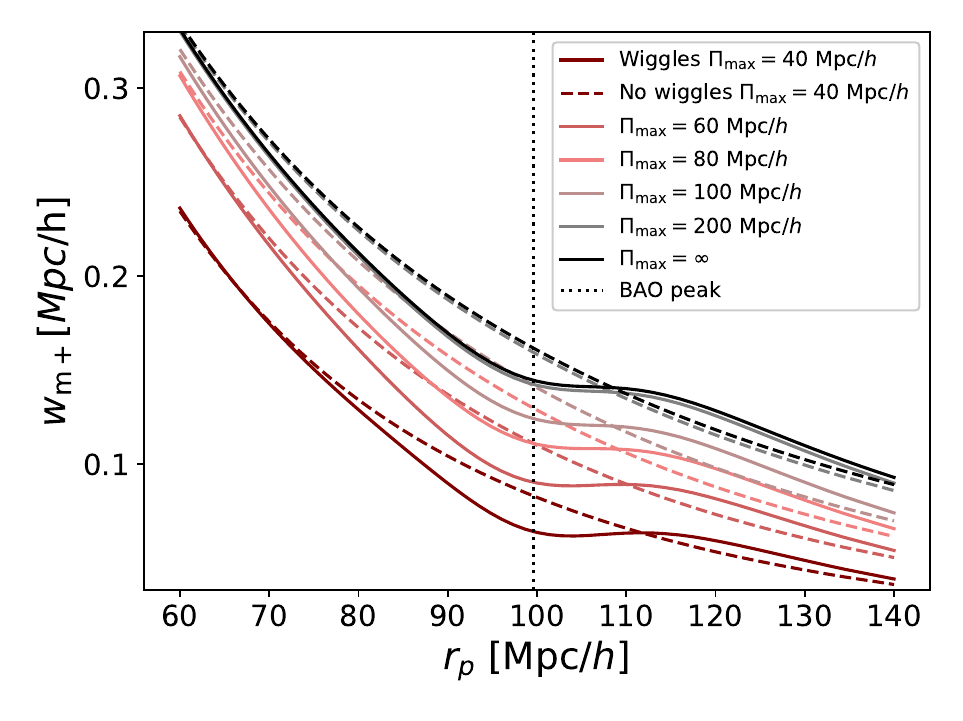}
    \caption{Projected correlation functions for matter clustering, $w_{\rm mm}$ (top) and alignments of galaxies with the matter field, $w_{\rm m+}$ (bottom), projected over different line-of-sight baselines, $\Pi_{\rm max}=[40,60,80]\,h^{-1}$ Mpc, for universes with (solid) and without (dashed) BAO. The BAO peak scale is indicated as a dotted vertical line. This corresponds to a peak in the case of $w_{\rm mm}$ and a trough for $w_{m+}$. }
    \label{fig:wproj}
\end{figure}

\section{Results}
\label{sec:results}

Figure \ref{fig:sketch} illustrates the geometry of the problem. Stacking on as many galaxies as possible and measuring the matter (or galaxy) distribution around them, one would find it slightly enhanced at scales equal to or smaller than the BAO comoving distance scale due to projection over the line-of-sight. The wider the range in $\Pi$, the higher the dilution of the BAO peak in projection, and the further in it will move in $r_p$. 

\subsection{Alignment correlations in short projection baselines}

Figure \ref{fig:wproj} shows the projected matter correlation function (top panel), computed at $z=0$, for different values of $\Pi_{\rm max}$ in a universe with and without wiggles. BAO features as an enhancement of the correlation function at a projected comoving separation of approximately $\sim 100$ $h^{-1}$ Mpc. Because of projection effects, for increasing projection baselines, such a distance is slightly reduced compared to the comoving distance at which one would find the BAO peak for the three-dimensional correlation function of matter. The larger the projection baseline ($\Pi_{\rm max}$), the further the peak moves towards smaller separations. 

In the bottom panel of Figure \ref{fig:wproj}, we show the projected alignment correlation function, computed at $z=0$, for different values of $\Pi_{\rm max}$ in a Universe with and without wiggles. Compared to $w_{\rm mm}$ in the top panel of Figure \ref{fig:wproj}, we see clearly that, at the location of the original BAO peak, there is now a trough, followed by a peak at a larger distance. This is indeed the feature that was seen in previous theoretical predictions and numerical simulations. 

To explain why it differs so from $w_{\rm mm}$, we make the following simplification of the problem: we assume that the BAO is a spherical shell centered at the origin, and that we are interested in computing the tides produced by this shell in the radial direction: $\tau_{rr}$ and in the $\theta$ direction: $\tau_{\theta\theta}$, according to Eqs. \ref{eq:taurr} and \ref{eq:tautt}, respectively. This can be combined to predict $\gamma^{\rm BAO}_+$. Our assumptions are justified by our findings in Figure \ref{fig:wproj}, in which we see the BAO feature appear as a peak in $w_{\rm mm}$ (top panel).

\begin{figure}[t]
    \centering
    \includegraphics[width=0.47\textwidth]{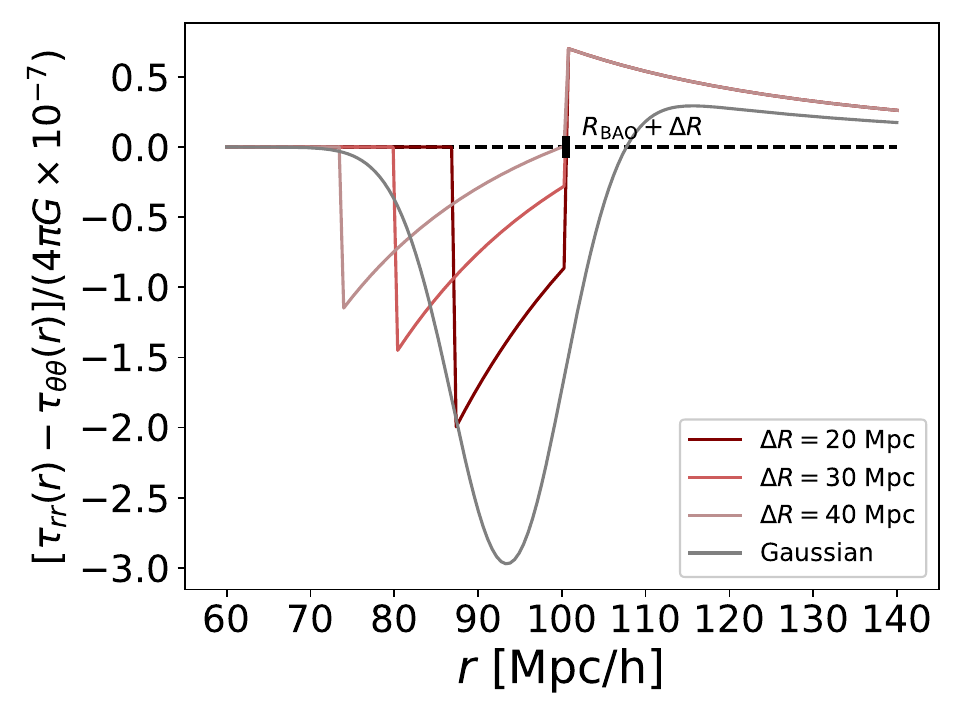}
    \caption{$\gamma_+^{\rm BAO}$ given the tidal field (normalized) of a spherical mass shell configuration spanning from $R_{\rm BAO}$ to $R_{\rm BAO}+\Delta R$ and assuming $C_1=1$ for illustration purposes. We plot the function for different BAO widths: $\Delta R=[20,30,40]$ Mpc in shades of red. In region III, where $r>R_{\rm BAO}+\Delta R$, the increase in the tidal field is consistent with the addition of a point mass $M_{\rm BAO}$. In Region II, within the BAO shell, we see a suppression of the tidal field compared to the `no wiggles' case. In Region I, inside the BAO shell, the tidal field remains unchanged. We also plot $\gamma_+^{\rm BAO}$ as originated from a spherical mass shell with a Gaussian profile centered at $R_{\rm BAO}+\Delta R/2$ and with a dispersion of $\Delta R/2$ (gray).}
    \label{fig:tides}
\end{figure}

$\gamma^{\rm BAO}_+$ is shown in Figure \ref{fig:tides}. This should be interpreted as the change in the intrinsic shapes of elliptical galaxies from a universe with BAO to a universe without BAO. (For illustration purposes, we adopt here $C_1=1$.) It is calculated by fixing the outer rim of the shell to $R_{\rm BAO}+\Delta R= 150$ Mpc and varying the choice of $\Delta R$. The overall mass normalization, $M_{\rm BAO}$ is arbitrary, but conserved, while varying $\Delta R$. Consistently with Eq. \ref{eq:taurr} we see that as a result of the BAO matter shell, the tidal field is unchanged inside the shell (Region I: $R<R_{\rm BAO}$), it decreases within the shell and it increases outside of it. The increase is originated by the addition of the mass $M_{\rm BAO}$, compared to the case where this is absent.

\begin{figure}[t]
    \centering
    \includegraphics[width=0.47\textwidth]{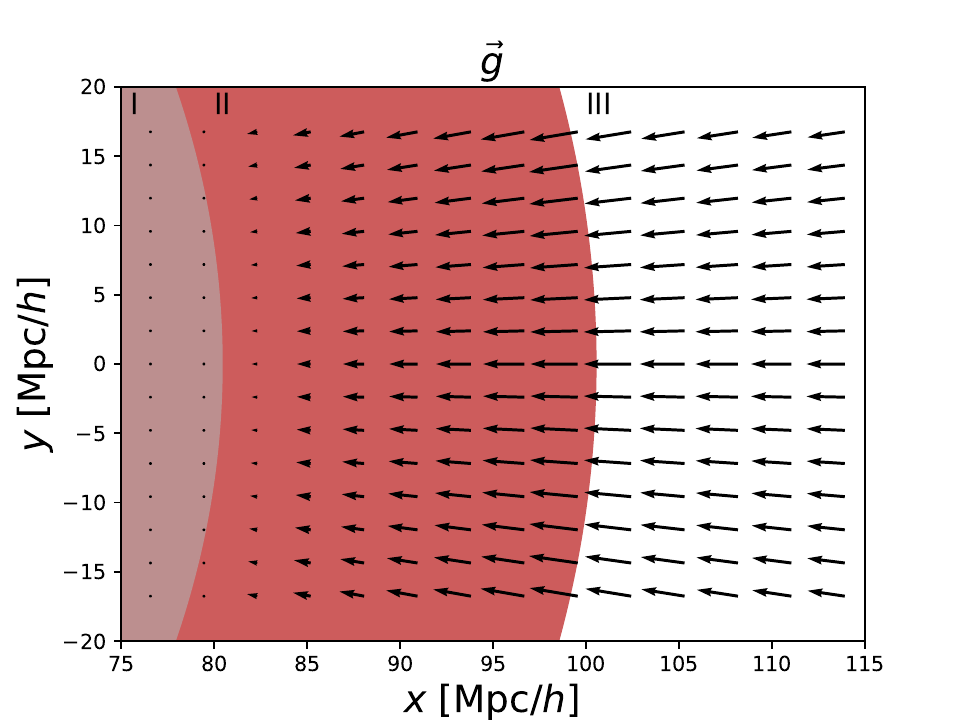}
    \caption{The gravitational acceleration vector, $\vec g$ from a spherical shell of matter. Three regions are indicated: region I inside the shell, region II within the shell and region III outside the shell. There is no gravity in region I. It builds up in region II and is the same as for a point mass with $M_{\rm BAO}$ in region III.}
    \label{fig:gvec}
\end{figure}

The Gaussian model (gray curve) represents a slightly more realistic situation in which the BAO has no sharp edge. For this case, we only show one possible scenario with a dispersion which corresponds to $10$ Mpc. The behaviour of the curve is similar in general to the hard-edge model, although $\gamma^{\rm BAO}_+$ transitions from negative to positive values at larger separations, above $R_{\rm BAO}+\Delta R$.

A model galaxy represented by a sphere embedded in this tidal field is deformed in the following way. Inside the shell, in region I, there is no deformation. Within the shell, in region II, the gravitational force increases with separation. This can be seen in the two-dimensional representation of the gravitational acceleration vector ($\vec g = -\nabla\phi$) shown in Figure \ref{fig:gvec}. The tidal field is thus negative in region II and thus compressive along the radial direction. Outside the shell, in region III, tidal forces are positive and thus disruptive, elongating the galaxy along the radial direction. This is due to the gravitational force decreasing outside the shell in the radial direction. 
Notice that the Fourier transform presented in Eq. \ref{eq:xig+LA} plays no role in yielding the trough feature. 
It is rather the fact that we are correlating positions with the tidal field that creates the feature, and this fact is encoded in the different $k$-factors and Bessel function inside Eq. \ref{eq:wg+LA}.

\begin{figure*}[t]
    \centering
    \includegraphics[width=0.47\textwidth]{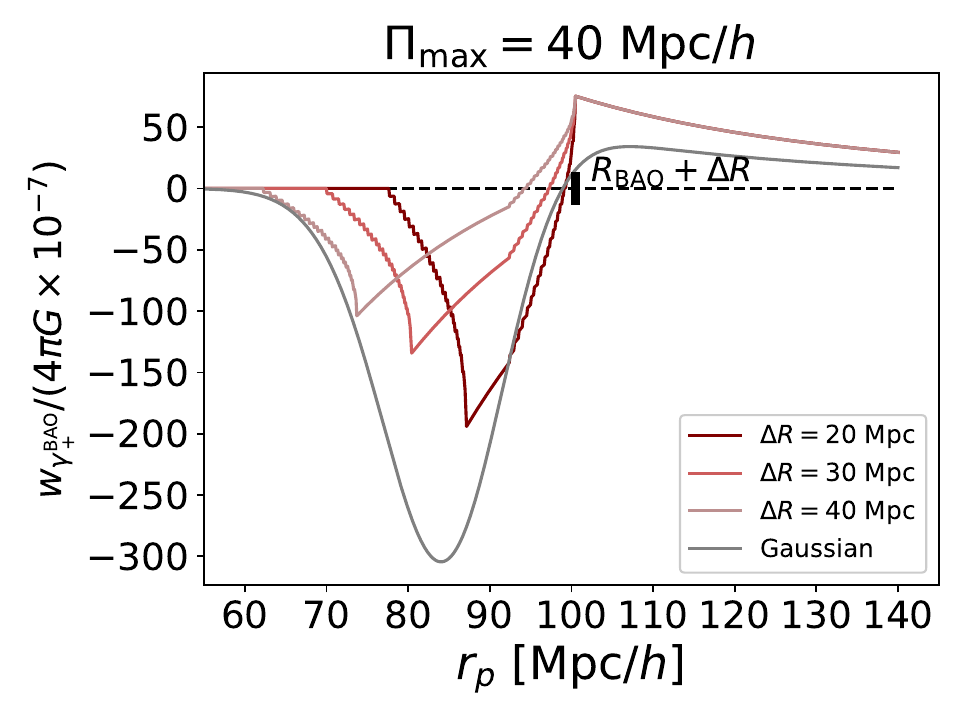}
    \includegraphics[width=0.47\textwidth]{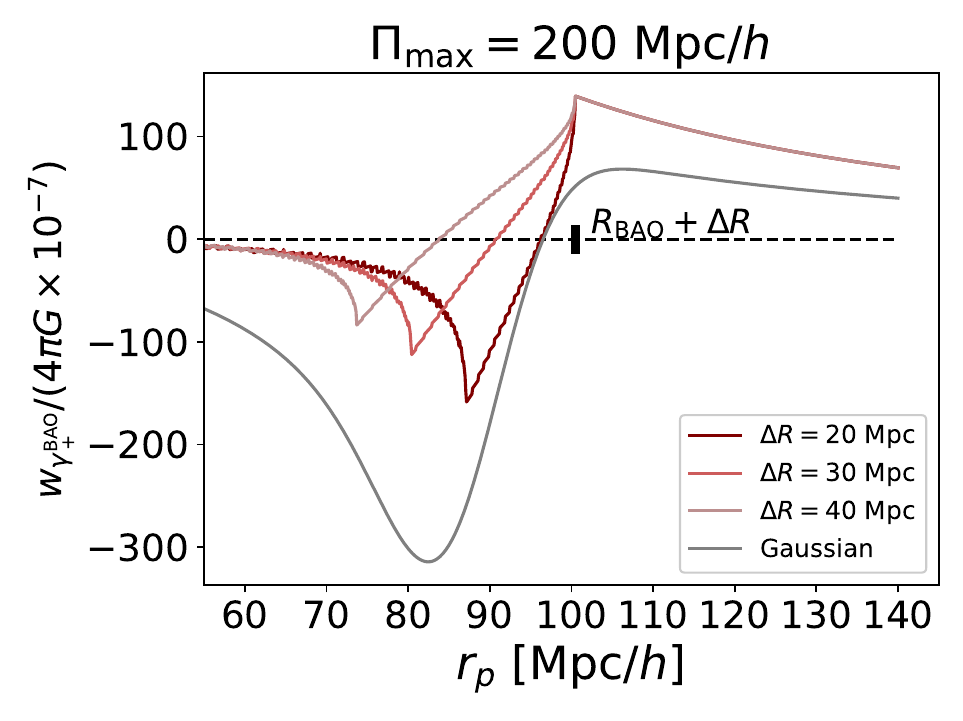}
    \caption{Projection over the line of sight of $\gamma^+_{\rm BAO}$. We plot the results for different BAO widths: $\Delta R=[20,30,40]$ Mpc in shades of red. The case of a Gaussian profile centered at $R_{\rm BAO}+\Delta R/2$ and with a dispersion of $\Delta R/2$ is also shown in gray. The left panel represents a short projection baseline with $\Pi_{\rm max}=40$ Mpc$/h$, and the right panel represents a longer baseline with $\Pi_{\rm max}=200$ Mpc$/h$. The qualitative shape of the BAO (a through followed by a peak) remains in both cases, but the details are sensitive to the projection length and the BAO model.}
    \label{fig:projgamma}
\end{figure*}

In Figure \ref{fig:projgamma}, we show the impact of projecting our model for $\gamma_+^{\rm BAO}$ over the line of sight. The qualitative form of the BAO feature remains the same in projection. For short projection baselines, the results are similar to the three-dimensional case, as expected. Increasing the projection baseline shifts the transition between trough and peak to smaller projected separations. The projection effect also enhances the differences between the model predictions when the projection baseline is longer.

\begin{figure}[t]
    \centering
    \includegraphics[width=0.47\textwidth]{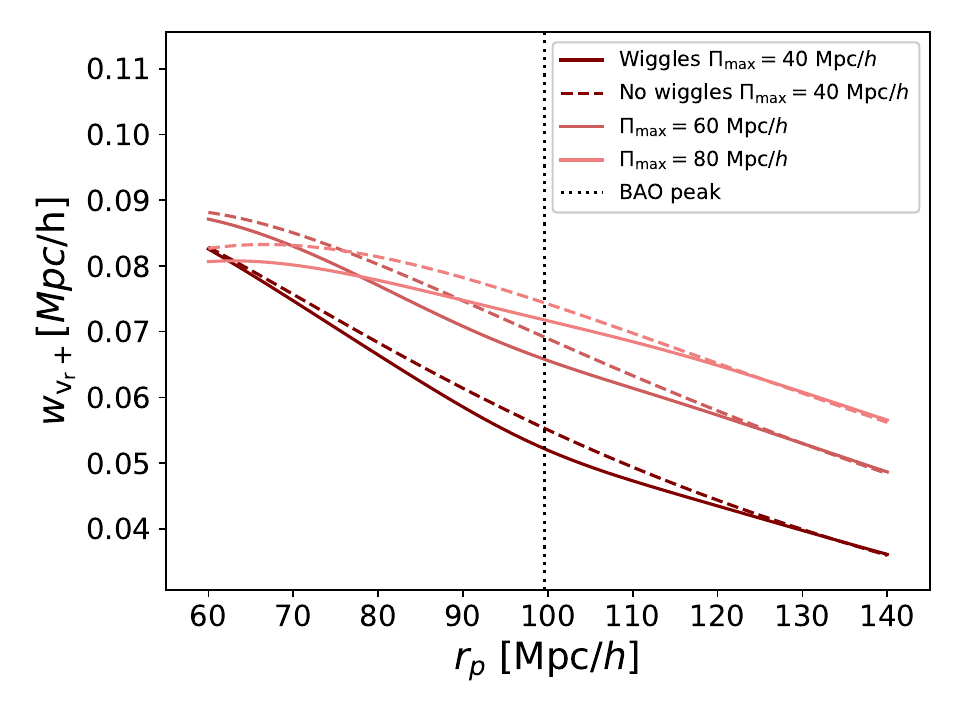}
    \caption{Projected correlation function for line-of-sight velocity-alignment statistics, $w_{v_r+}$, projected over different line-of-sight baselines, $\Pi_{\rm max}=[40,60,80]\,h^{-1}$ Mpc, for universes with (solid) and without (dashed) BAO. The BAO peak scale is indicated as a dotted vertical line. This corresponds to a trough in $w_{v_r+}$.}
    \label{fig:wprojv}
\end{figure}

We also obtained the line-of-sight velocity-intrinsic shape projected correlation function, shown in Figure \ref{fig:wprojv}. This shows very similar BAO behaviour to $w_{\rm m+}$ in the bottom panel of Figure \ref{fig:wproj}. There is a trough at the BAO scale, followed by an excess at larger scales compared to the `no wiggles' case. This is justified by the fact that at these scales, the velocity field of the large-scale structure follows the linear continuity equation, resulting in $v(k) \propto \delta(k)/k$. It is thus not surprising that the BAO would also follow qualitatively the tidal field of the spherical shell of mass as presented in Figure \ref{fig:tides}, confirming the findings of \citep{Okumura19}.

\begin{figure}
    \centering
    \includegraphics[width=0.47\textwidth]{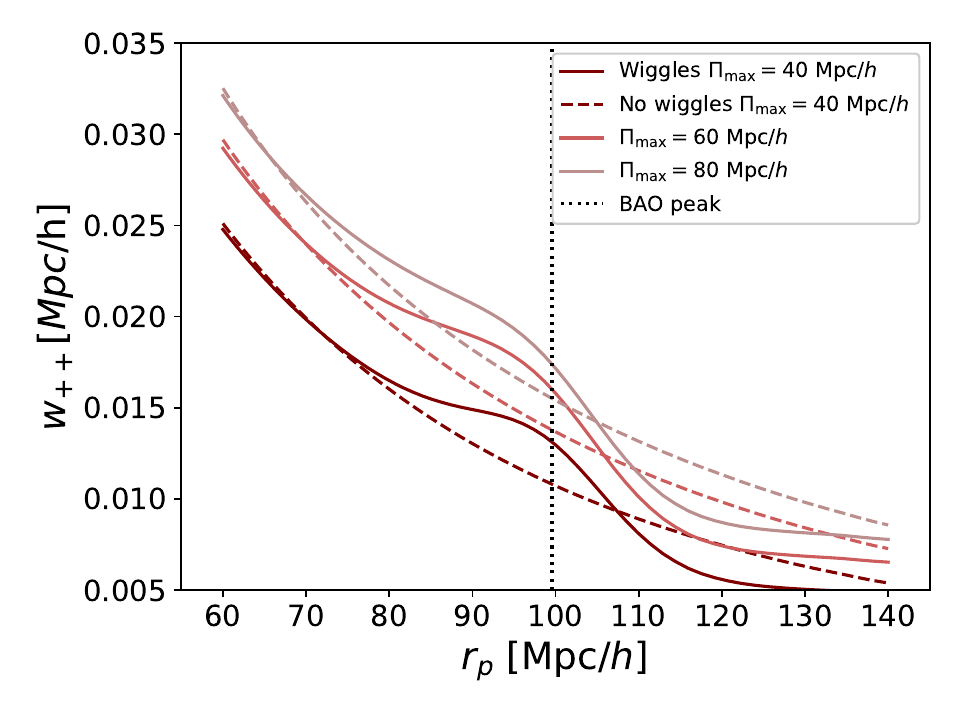}
    \includegraphics[width=0.47\textwidth]{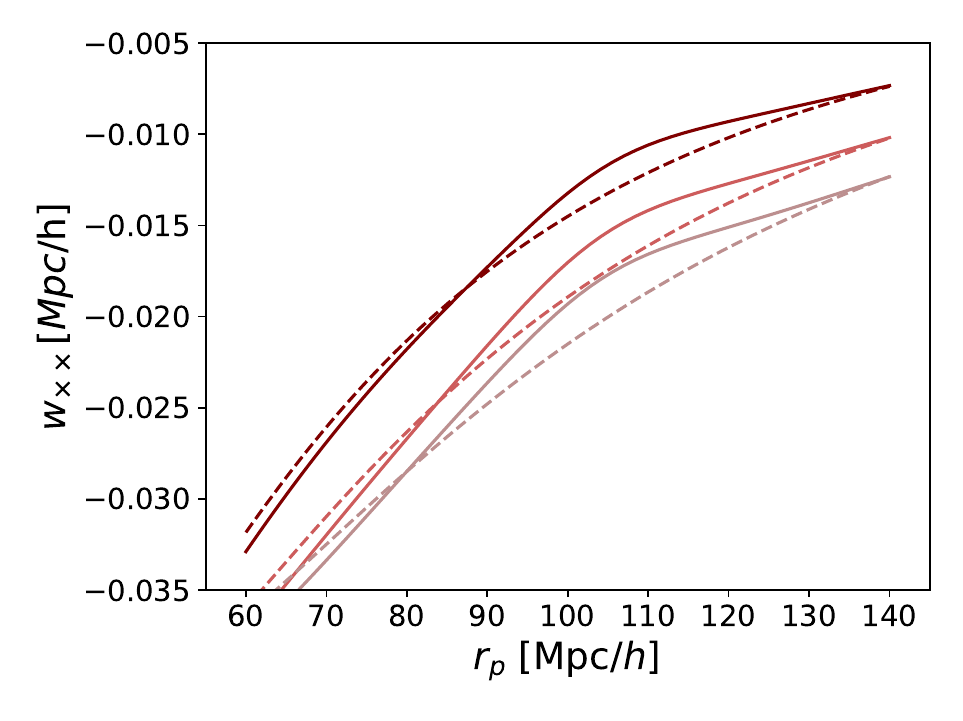}
    \caption{Projected shape-shape correlation functions for $++$ (top) and $\times\times$ (bottom), projected over different line-of-sight baselines, $\Pi_{\rm max}=[40,60,80]\,h^{-1}$ Mpc, for universes with (solid) and without (dashed) BAO. The BAO peak scale is indicated as a dotted vertical line.}
    \label{fig:wss}
\end{figure}

For completion, we also show in Figure \ref{fig:wss} the impact of the BAO feature in shape-shape correlations. BAO appear as a peak in the $w_{\rm ++}$ correlation function rather than a trough. This can be interpreted as the consequence of two sign cancellations in the product of $\gamma_+^{\rm BAO}$. Our toy model gives null predictions for the $\times$ component and is not useful here. But we see in the bottom panel of Figure \ref{fig:wss} that the linear alignment model predicts that the BAO feature in $w_{\times\times}$ appears as a peak at scales larger than the BAO scale. In practice, little signal-to-noise is expected in this particular correlation \citep{Blazek11}. 


\begin{figure}
    \centering
    \includegraphics[width=0.47\textwidth]{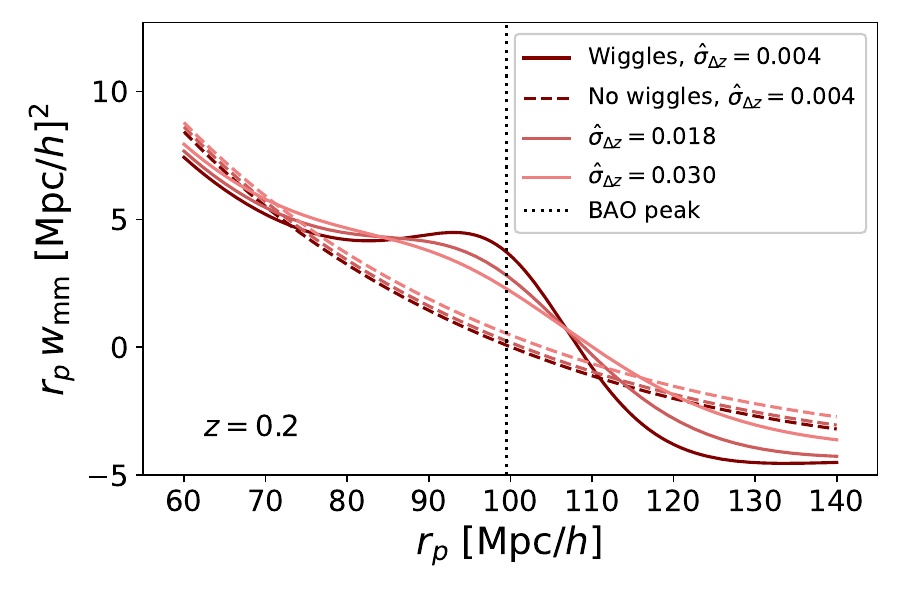}
    \includegraphics[width=0.47\textwidth]{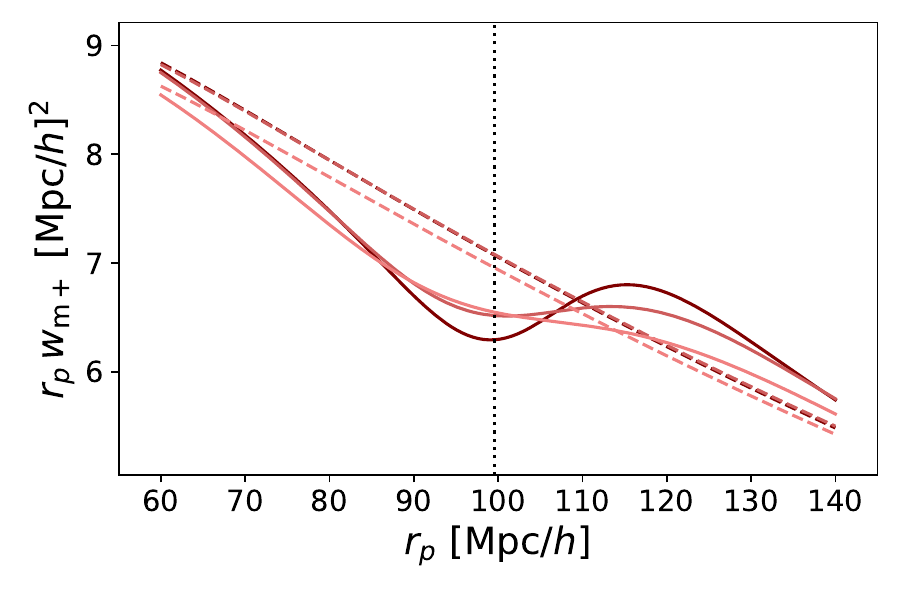}
    \includegraphics[width=0.47\textwidth]{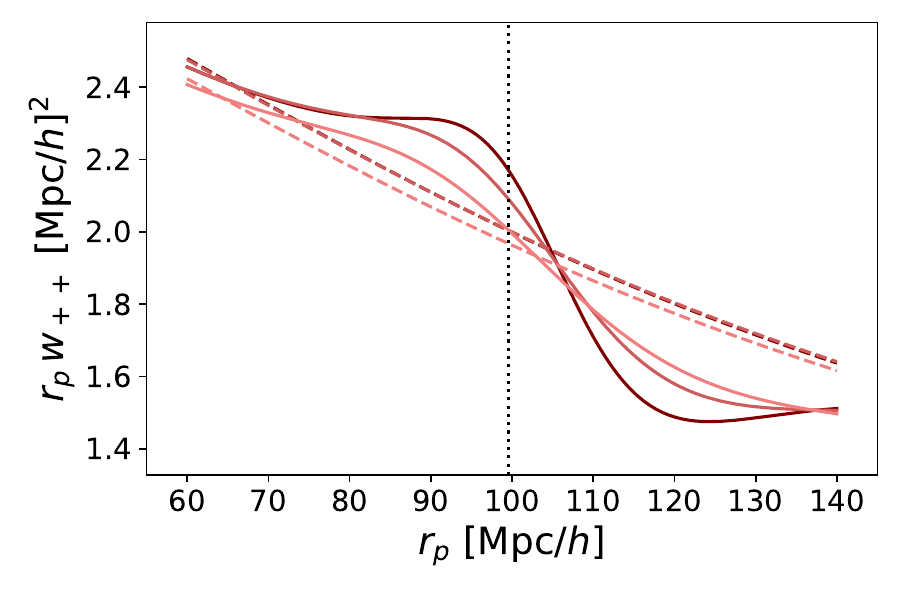}
    \caption{Projected correlation functions for matter-matter (top), matter-shape (middle) and shape-shape (bottom) correlations computed in the case of redshift uncertainties, quantified by $\hat{\sigma}_{\Delta z} = [0.004, 0.018, 0.3]$, for universes with (solid) and without (dashed) BAO. The BAO peak scale is indicated as a dotted vertical line.}
    \label{fig:wab_phot}
\end{figure}

\begin{figure}
    \centering
    \includegraphics[width=0.47\textwidth]{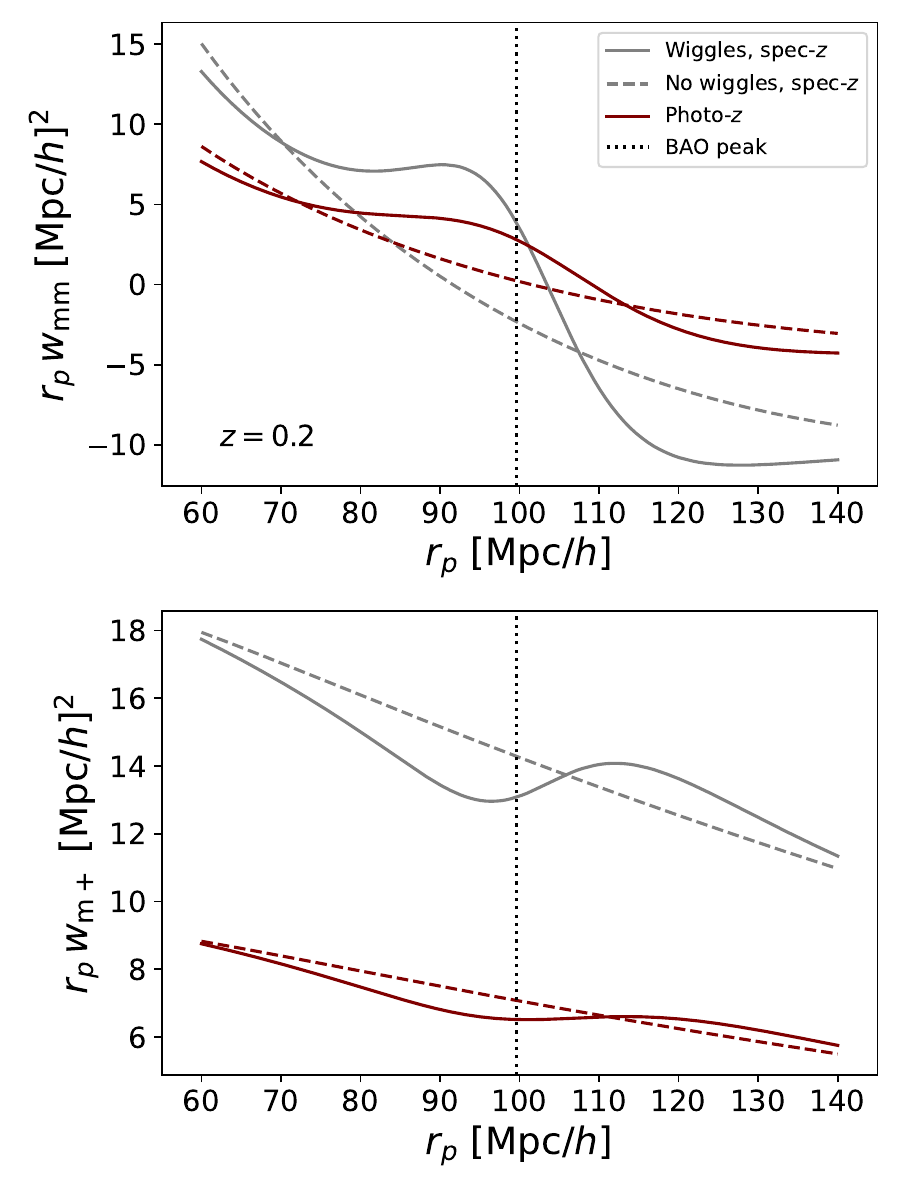}
    \caption{Projected correlation functions for matter-matter (top) and matter-shape (bottom) correlations computed in the case of accurate, spectroscopic (grey) and photometric (maroon) redshift information, for universes with (solid) and without (dashed) BAO. The BAO peak scale is indicated as a dotted vertical line.}
    \label{fig:wab_specz_photoz}
\end{figure}

\subsection{Long projection baselines and photometric redshifts}

The bottom panel of Figure \ref{fig:wproj} presents $w_{\rm m+}$ integrated over an infinite projection baseline. We see that as $\Pi_{\rm max}$ increases, BAO become progressively smeared. The evolution of the amplitude of $w_{\rm m+}$ is monotonic and information progressively saturates as $\Pi_{\rm max}\rightarrow \infty$. In practice, most observational works adopt $60\,h^{-1}\,{\rm Mpc}<\Pi_{\rm max}<100\,h^{-1}\,{\rm Mpc}$. The shape of the BAO is preserved with increasing $\Pi_{\rm max}$. This is similar in the case of $w_{\rm mm}$ in the top panel of Figure \ref{fig:wproj}, though here the correlation amplitude does not change monotonically. For all cases, we notice the BAO feature (peak or trough depending on the fields considered) move inwards as a consequence of the increased projection length.

Next, we address the impact of redshift uncertainty, such as in the case of photometrically obtained redshift information (photo-$z$), on the projected correlation functions. We choose three different uncertainty scenarios: redshifts obtain by narrow-band photometry with $\hat{\sigma}_{\Delta z}\sim0.004$ (e.g. a survey such as PAUS \cite{PAUSPZ}), redshifts obtained over a bright galaxy sample or using the galaxy red sequence with $\hat{\sigma}_{\Delta z}\sim0.018$ \cite{redmagic, KiDSLRG, KiDSBrightPZ} and redshifts obtained from an optimized \emph{gold} sample from large photometric surveys with $\hat{\sigma}_{\Delta z}\sim0.03$ (e.g. for surveys such as KiDS or DES \cite{KiDSSOMPZ, DESBAOPZ}). The last value is also equal to the assumed value of the photo-z scatter for large-scale structure tracers in next generation photometric surveys, such as the Vera Rubin Observatory LSST \cite{DESCrequirements}.

Figure \ref{fig:wab_phot} shows the projected correlation functions $w_\mathrm{mm}, w_\mathrm{m+}$ and $w_{++}$ computed at $z=0.2$, in a universe with and without BAO, for the three different redshift uncertainty scenarios. We see that, as the uncertainty gets larger, the BAO feature is less pronounced for all three functions. The behaviour of the clustering and alignment signals are similar to the case with accurate, spectroscopic redshifts (spec-$z$). The choice of $z=0.2$ is explicitly made to avoid negative values of redshift induced by the SMAD scatter model.

It is also interesting to compare the projected correlation functions in the case of no redshift uncertainty and an infinite $\Pi_\mathrm{max}$ to functions with modelled redshift uncertainty. We show this in Figure \ref{fig:wab_specz_photoz} where the photo-$z$ signal has $\hat{\sigma}_{\Delta z}\sim0.018$. The signal obtained through photo-$z$'s is closer to zero in both the clustering and alignment correlation. Since the clustering signal crosses zero at around 110 Mpc/$h$, the photometric clustering signal appears simply flatter. In the case of matter-shape correlations, the photo-$z$ signal is about a factor of 2 smaller than the spec-$z$.

\section{Conclusion}
\label{sec:conclusions}

While BAO appear as a peak in the matter field projected auto-correlation, in the correlation of matter with intrinsic galaxy shapes, the pattern is replaced by a trough at the same scale, followed by an excess at larger separations. We showed that this behavior is consistent with the response of galaxy shapes to the linear tidal field represented by a shell of matter with radius similar to the location of the BAO peak. A similar behavior is observed for the correlation between intrinsic shapes and radial velocities. 

Our work highlights the need for dedicated templates for the BAO in such statistic, if a detection is to be attempted. This is, in fact, not far from the reach of current surveys \citep{CD13}. While our pedagogical model based on a matter shell successfully describes the BAO feature at a qualitative level, more accurate fits to the data can be obtained by using the matter power spectrum directly.

Progressively increasing projection baselines for the correlation function results in a smearing of the BAO peak. When measuring this from data, increasing the value of $\Pi_\mathrm{max}$ increases the number of uncorrelated pairs used in estimating the projected correlation functions which results in loss of signal-to-noise. Therefore, there exists a value of $\Pi_\mathrm{max}$ that maximises the signal-to-noise and typically values around 60 $h^{-1}$ Mpc have been chosen in previous studies, when working with spectroscopic redshifts.

In the case of redshift uncertainty, such as for a sample where the redshift was obtained through photometry, two effects take place in the projected correlation functions. Firstly, the correlation function is closer to zero and the signal is lower. The second effect is that the BAO feature is washed out by the redshift uncertainty. The higher the uncertainty, the less pronounced the BAO feature will be, across all correlation functions. 

We note that photometric data are easier to obtain compared to spectroscopic and, as a consequence, photometric samples are typically much larger than spectroscopic ones. This increase in sample size can result in a higher signal-to-noise obtained for correlation functions in the case of samples with redshift uncertainty even though the signal itself is lower. A more quantitative analysis of this, together with forecasts for upcoming surveys, is left for future work.

\begin{acknowledgments}
This publication is part of the project ``A rising tide: Galaxy intrinsic alignments as a new probe of cosmology and galaxy evolution'' (with project number VI.Vidi.203.011) of the Talent programme Vidi which is (partly) financed by the Dutch Research Council (NWO). This work is also part of the Delta ITP consortium, a program of the Netherlands Organisation for Scientific Research (NWO) that is funded by the Dutch Ministry of Education, Culture and Science (OCW).
\end{acknowledgments}

\newpage

\appendix

\begin{widetext}

\section{Limber approximation}

In this appendix we explicitly show the derivation of Eq. \ref{eq:wm+_approx} by making use of the Limber approximation \citep{Limber}. For completeness, we will consider the $\rm g+$ correlation instead of $\rm m+$ and we will explicitly model the window functions for the galaxy populations used to trace the density and shape fields. These will be labelled $q_{\rm g}(\chi)=dN_{\rm g}/d\chi$ and $q_\gamma(\chi)=dN_\gamma/d\chi$ for number and shape tracers, respectively, and where $\chi$ is the comoving line-of-sight distance.

First, we establish that our goal is to calculate Eq. \ref{eq:wg+} to the case where $\Pi_{\rm max}\rightarrow \infty$:
\begin{equation}
    w_{\rm g+}(r_p) = \int d\chi  \,q_{\rm g}(\chi)\int d\chi'\,q_\gamma(\chi') \langle \delta(\vec x_p,\chi)\gamma_+(\vec x_p',\chi')\rangle.
\end{equation}
The Limber approximation \citep{Limber} consists of assuming that the galaxy positions and the intrinsic $+$ component of the shape field are uncorrelated unless they are evaluated at the same redshift or line-of-sight distance. In other words, there is a coherence scale \citep{Bartelmann} over which the correlation is non-zero and this is much smaller than the infinite projection baseline we are using to project $\xi_{\rm m+}$. 

Replacing the three-dimensional correlation function by its Fourier transform, we obtain
\begin{equation}
    w_{\rm g+}(r_p) = \int d\chi  \,q_{\rm g}(\chi)
    \int d\chi'\,q_\gamma(\chi') 
    \int \frac{d^3k}{(2\pi)^3}
    \int \frac{d^3k'}{(2\pi)^3}
    \langle \hat\delta(\vec k,\chi)
    \hat\gamma_+(\vec k',\chi')\rangle 
    e^{-i\vec k_\perp \cdot x_\perp}
    e^{-i\vec k_\perp' \cdot x_\perp'}
    e^{-ik_z\chi}e^{-ik_z'\Pi'}.
\end{equation}
Here we have aligned the component of the wavevector that is perpendicular to the line-of-sight with the $x$ axis without loss of generality \citep{Blazek11}. 
By explicitly modelling the power spectrum of the density and the shapes, we can write
\begin{equation}
    w_{\rm g+}(r_p) = \int d\chi  \,q_{\rm g}(\chi)\int d\chi'\,q_\gamma(\chi') \int \frac{d^3k}{(2\pi)^3}\int \frac{d^3k'}{(2\pi)^3}P_{\rm m+}(\vec k,z) (2\pi)^3\delta_D(\vec k-\vec k')    e^{-i\vec k_\perp \cdot x_\perp}
    e^{-i\vec k_\perp' \cdot x_\perp'}
    e^{-ik_z\chi}e^{-ik_z'\Pi'}.
\end{equation}
and collapse one of the integrals in wavevector to obtain
\begin{equation}
    w_{\rm g+}(r_p) = \int d\chi  \,q_{\rm g}(\chi)\int d\chi'\,q_\gamma(\chi') \int \frac{d^3k}{(2\pi)^3}P_{\rm m+}(\vec k,z) e^{-i\vec k_\perp\cdot(\vec x_\perp-\vec x_\perp')}e^{-ik_z\chi}e^{-ik_z\chi'}.
\end{equation}
Applying the Limber approximation,
\begin{equation}
    w_{\rm g+}(r_p) = \int d\chi  \,q_{\rm g}(\chi)\,q_\gamma(\chi) \int d\chi' \int \frac{d^3k}{(2\pi)^3}P_{\rm m+}(\vec k,z) e^{-i\vec k_\perp\cdot(\vec x_\perp-\vec x_\perp')}e^{-ik_z\chi} e^{-ik_z\chi'}.
\end{equation}
From here onward, we will assume $q_{\rm g}(\chi)=q_\gamma(\chi)=\delta_D(\chi)$, which corresponds to correlations are evaluated at $z=0$ for simplicity.
The integral over $\chi'$ can now be brought inside, resulting in a Dirac delta over the line-of-sight wavevector:
\begin{equation}
    w_{\rm g+}(r_p) =  \int \frac{d^3k}{(2\pi)^3}P_{\rm g+}(\vec k,z=0) e^{-i\vec k_\perp\cdot(\vec x_\perp-\vec x_\perp')}e^{-ik_z\chi}2\pi\delta_D(k_z).
\end{equation}
Before continuing we re-write $P_{\rm g+}$ explicitly:
\begin{equation}
    w_{\rm g+}(r_p) = - \frac{C_1\rho_{\rm crit}\Omega_{\rm m}b_{\rm g}}{D(z)}\int \frac{d^3k}{(2\pi)^3}P(k,z=0)\frac{k_x^2-k_y^2}{k^2} e^{-i\vec k_\perp\cdot(\vec x_\perp-\vec x_\perp')}e^{-ik_z\chi}2\pi\delta_D(k_z).
\end{equation}
where $\vec k_\perp=(k_x,k_y)$. The presence of the Dirac delta in $k_z$ simplifies the whole expression to
\begin{equation}
    w_{\rm g+}(r_p) = -\frac{C_1\rho_{\rm crit}\Omega_{\rm m}b_{\rm g}}{D(z)} \int \frac{d^2k_\perp}{(2\pi)^2}P(k_\perp,z=0)\frac{k_x^2-k_y^2}{k_\perp^2} e^{-i\vec k_\perp\cdot(\vec x_\perp-\vec x_\perp')}.
\end{equation}
If $\theta_k$ is the angle between $\vec k_\perp$ and the $x$ axis and $k_\perp=|\vec k_\perp|$, then
\begin{equation}
    w_{\rm g+}(r_p) = - \frac{C_1\rho_{\rm crit}\Omega_{\rm m}b_{\rm g}}{D(z)} \int \frac{dk_\perp d\theta_k\,k_\perp}{(2\pi)^2}P(k_\perp,z=0)\cos(2\theta_k) e^{-ik_\perp r_p \cos\theta_k}.
\end{equation}
This makes the second order Bessel function appear and now the integral is over the absolute value of 
\begin{equation}
    w_{\rm g+}(r_p) = \frac{C_1\rho_{\rm crit}\Omega_{\rm m}b_{\rm g}}{D(z)} \int \frac{dk_\perp}{2\pi}k_\perp P(k_\perp,z)J_2(k_\perp r_p).
\end{equation}
Similarly for the correlation with the matter field,
\begin{equation}
    w_{\rm m+}(r_p) = \frac{C_1\rho_{\rm crit}\Omega_{\rm m}}{D(z)} \int \frac{dk_\perp}{2\pi}k_\perp P(k_\perp,z)J_2(k_\perp r_p).
\end{equation}
This is in agreement with Eq. \ref{eq:wm+_approx}.

\end{widetext}

\bibliography{biblio}

\end{document}